\documentclass[11pt]{article}

\usepackage[preprint]{acl}
\usepackage{times}
\usepackage{latexsym}
\usepackage{microtype}
\usepackage{inconsolata}
\usepackage{graphicx}
\usepackage{color, colortbl}
\usepackage[table]{xcolor}
\usepackage{url}
\usepackage{tabularx}
\usepackage{subcaption}
\usepackage{textcomp}
\usepackage{soul}
\usepackage{multirow}
\usepackage{enumitem}
\usepackage{mathtools}
\usepackage{siunitx}

\usepackage{booktabs} 
\usepackage{longtable,booktabs}

\usepackage{arydshln}
\definecolor{linkColor}{RGB}{6,125,233}
\definecolor{green}{rgb}{0.0, 0.65, 0.31}
\definecolor{bleudefrance}{rgb}{0.19, 0.55, 0.91}
\definecolor{ceruleanblue}{rgb}{0.16, 0.32, 0.75}
\definecolor{grey}{HTML}{969696}
\definecolor{violet}{HTML}{756bb1}
\definecolor{dgrey}{HTML}{01665e}
\definecolor{lgrey}{HTML}{5ab4ac}
\definecolor{dgreen}{HTML}{005a32}
\definecolor{purple}{HTML}{ae017e}

\definecolor{editCol}{HTML}{000000}
\definecolor{maskCol}{HTML}{c51b7d}
\definecolor{lrColor}{HTML}{8856a7}
\definecolor{trColor}{HTML}{d01c8b}
\definecolor{ctColor}{HTML}{4dac26}
\definecolor{brickred}{HTML}{f03b20}

\definecolor{sigone}{RGB}{222, 235, 247}    
\definecolor{sigtwo}{RGB}{158, 202, 225}    
\definecolor{sigthree}{RGB}{107, 174, 214}  

\definecolor{improveCol}{HTML}{4dac26}
\definecolor{worsenCol}{HTML}{d01c8b}

\definecolor{DarkBlue}{HTML}{00008B}
\definecolor{mscolor}{HTML}{01665e}
\definecolor{nmscolor}{HTML}{bf812d}
\definecolor{lgreen}{HTML}{ccece6}
\definecolor{dolive}{HTML}{308014}

\colorlet{tablerowcolor4}{gray!50} 

\newcommand*{\textlabel}[2]{%
  \edef\@currentlabel{#1}
  \phantomsection
  #1\label{#2}
}

\colorlet{tableheadcolor}{gray!25} 
\colorlet{tablerowcolor}{gray!10} 
\colorlet{tablerowcolor2}{gray!45} 
\colorlet{tablerowcolor3}{gray!12} 

\newcommand{\rowcollight}{\rowcolor{tablerowcolor3}} %

\newcolumntype{a}{>{\columncolor{tablerowcolor}}r}
\definecolor{aicolor}{HTML}{018571}
\definecolor{occolor}{HTML}{ff7799}

\definecolor{aicolor}{HTML}{fc8d62}
\definecolor{occolor}{HTML}{253494}

\newif{\ifhidecomments}
  \hidecommentsfalse 
\ifhidecomments
    \newcommand{\olivia}[1]{}
    \newcommand{\yunhao}[1]{}
    \newcommand{\talayeh}[1]{}
    \newcommand{\koustuv}[1]{}
\else
    \newcommand{\olivia}[1]{\textbf{\small\sffamily{\textcolor{DarkBlue}{[#1 -- Olivia]}}}}
    \newcommand{\yunhao}[1]{\textbf{\small\sffamily{\textcolor{dgreen}{[#1 -- Yunhao]}}}}
    \newcommand{\talayeh}[1]{\textbf{\small\sffamily{\textcolor{orange}{[#1 -- Talayeh]}}}}
    \newcommand{\koustuv}[1]{\textbf{\small\sffamily{\textcolor{purple}{[#1 -- Koustuv]}}}}
  \fi

\colorlet{tableheadcolor}{gray!25} 

\definecolor{neutralCol}{HTML}{dd1c77}
\definecolor{neutralGreen}{HTML}{31a354}
\definecolor{NewBlue}{HTML}{1879ba}
\definecolor{bleudefrance}{rgb}{0.19, 0.55, 0.91}  
\definecolor{AfTrColor}{HTML}{0868ac}  
\definecolor{BfTrColor}{HTML}{a8ddb5}  

\definecolor{AfCtColor}{HTML}{b10026}  
\definecolor{BfCtColor}{HTML}{fd8d3c}

\graphicspath{ {figures/} }

\newcommand{\sechead}[1]{%
  \multicolumn{14}{>{\columncolor{tablerowcolor3}}l}{\textit{#1}}}
  
\newcommand{\para}[1]{\vspace{0.2em}\noindent\textbf{\textit{#1}~}}

\usepackage{graphicx}

\title{X-pressing Intimacy: Mental Health Disclosure After Twitter Acquisition}
\author{
    \textbf{Olivia Pal}\textsuperscript{*1},~
    \textbf{Yunhao Yuan}\textsuperscript{*2},~
    \textbf{Meghana Mandava}\textsuperscript{1},~
    \textbf{Talayeh Aledavood}\textsuperscript{2},\\
    \textbf{Eshwar Chandrasekharan}\textsuperscript{1},~
    \textbf{Koustuv Saha}\textsuperscript{1}\\
    \textsuperscript{1}University of Illinois Urbana-Champaign, Urbana, IL, USA\\
    \textsuperscript{2}Aalto University, Espoo, Finland\\
    \texttt{\{opal2, mmand13, eshwar, ksaha2\}@illinois.edu}\\
    \texttt{\{yunhao.yuan, talayeh.aledavood\}@aalto.fi}
}

\begin{document}


\maketitle
\begingroup
\renewcommand{\thefootnote}{\fnsymbol{footnote}}
\footnotetext[1]{Both authors contributed equally.}
\endgroup
\begin{abstract}
Social media platforms provide spaces for individuals to disclose mental health experiences and seek support, yet how these disclosures evolve during major shifts in platform governance remains unclear. We examine mental health self-disclosure surrounding Elon Musk's October 2022 acquisition of Twitter, analyzing approximately 824K tweets across anxiety, depression, bipolar disorder (BD), borderline personality disorder (BPD), and suicidal ideation (SI) or related experiences, alongside 266K comparison tweets. Using Linguistic Inquiry and Word Count (LIWC) and an LLM-based intimacy classifier, we characterize changes in psycholinguistic expression and disclosure intimacy. Following the acquisition, we observe declines in affective and social language, alongside increases in first-person singular pronouns and future-focused expressions. Intimacy patterns vary by condition: anxiety, depression, and BPD show increases in low-intimacy and decreases in moderate-intimacy disclosures, whereas suicide-related posts show more high-intimacy disclosures. However, compared with broader Twitter discourse, only the relative decline in moderate- versus low-intimacy disclosure for BPD remains statistically significant. Our findings underscore the importance of considering both language and intimacy when studying platform transitions, informing the design of online environments that support sensitive self-disclosure.

\end{abstract}

\section{Introduction~\label{section:intro}}

\begin{quote}
\small
``\textit{New Twitter policy is freedom of speech, but not freedom of reach}''---Elon Musk, after Twitter takeover~\cite{ordonez2022twitter}, announcing a moderation approach emphasizing reduced visibility of harmful content rather than its removal. 
\end{quote}

Social media platforms are spaces in which users build communities, establish safe spaces, and find support among others with relatable experiences~\cite{lucero2017safe,de2014mental,berry2017whywetweetmh}.
These platforms have long provided opportunities to share 
personal experiences with mental health, including experiences surrounding depression, anxiety, stress, suicidal ideation, stigma, treatment, and everyday challenges, while also seeking support from others with similar experiences~\cite{de2014mental,berry2017whywetweetmh,luo2020self,mendu2020framework,yuan2023minority}.
Users disclose such experiences on public platforms and may obtain coping, social support, and therapeutic benefits from doing so~\cite{de2014mental,berry2017whywetweetmh,yuan2023mental}.
The intimacy of these self-disclosures---the depth of personal information shared---is important because opening up about vulnerable experiences can help individuals build supportive relationships and potentially derive therapeutic benefits from self-disclosure on social media~\cite{ernala2017linguistic,ernala2018characterizing}.


That said, mental health disclosure depends not only on what individuals wish to share, but also on the social and platform environment in which these conversations take place.
Social Penetration Theory helps explain this dependence on context: the depth of self-disclosure is shaped by the perceived rewards and costs of revealing personal information~\cite{altman1973social}.
Audience expectations, community norms, platform affordances, and moderation practices can shape whether and how individuals disclose personal experiences online~\cite{ernala2018characterizing,marwick2011tweet,litt2016imagined,wise2006moderation,fiesler2018reddit,budenz2022if}.
For instance,~\citeauthor{ernala2018characterizing} showed that audience engagement predicts subsequent changes in disclosure intimacy. 


Consequently, when platform governance changes, the conditions under which users participate and share may change as well~\cite{chandrasekharan2017you,chandrasekharan2022quarantined}.
Changes in platform rules and moderation can reshape community norms, participation, and users' relationship with the spaces~\cite{wise2006moderation,fiesler2018reddit,southerton2021restricted}.
Such changes can also lead users to contest platform governance or migrate elsewhere~\cite{sybert2022demise,edwards2021migration}.
For instance, when Tumblr banned adult content in 2018, users contested the platform owners' authority and values surrounding the governance of the platform~\cite{sybert2022demise}, while the policy disrupted communities that had relied on Tumblr for marginalized and LGBTQ+ forms of expression~\cite{southerton2021restricted,lucero2017safe}.
Comparably, on October 27, 2022, Elon Musk acquired Twitter, initiating another substantial transformation in the governance of a major social media platform~\cite{murthy2024sociology,hickey2023auditing}.

Following the acquisition, Twitter reinstated previously suspended accounts and dissolved its Trust and Safety Council~\cite{reuters2022amnesty,dang2022twitter}.
Prior work revealed changes in hateful content~\cite{hickey2023auditing,hickey2025x}, participation among particular communities~\cite{chang2023environmental,bisbee2025vibes}, sentiment surrounding the acquisition~\cite{schmidt2023polarity}, and migration toward alternative social media platforms~\cite{cava2023drivers,zia2023flocking,jeong2024exploring}.
These findings show that the acquisition was accompanied by changes in both the content and communities.
However, the shifts in mental health disclosure remain underexplored. 
In particular, changes in platform governance may also have less visible costs: content that is never written. 
Users anticipating stigma, harassment, or unsupportive responses may withhold personal experiences or share them less intimately, limiting opportunities for connection and support. 

Accordingly, guided by Social Penetration Theory, we examine how the depth of sensitive self-disclosure varies across a platform governance transition,  
surrounding Twitter's acquisition.
We target the following research questions (RQs):

\para{\textbf{RQ1:}} How do symptomatic and psycholinguistic expressions of mental health change across a platform governance transition?




\para{RQ2:} How does the intimacy of mental health disclosures change across this transition?


To address our RQs, we compiled mental health discourse on Twitter surrounding the October 27, 2022 acquisition, focusing on self-disclosures of anxiety, depression, bipolar disorder (BD), borderline personality disorder (BPD), and suicidal ideation (SI) or related experiences.
We examine Twitter posts before and after the acquisition to characterize how mental health expression changed across this platform transition.
For our psycholinguistic analysis, we compare these changes against a comparable control group of Twitter posts to distinguish shifts specific to mental health discourse from broader changes in language on Twitter.
We examine disclosure intimacy using a three-level measure ranging from less personal references to highly intimate accounts of lived mental health experiences.

Our findings reveal changes in both the language and intimacy of mental health self-disclosure following Twitter's acquisition. 
Mental health Twitter posts exhibited declines in affective language, particularly anxiety and sadness, and in social references, alongside increases in first-person singular pronouns and future-focused language. 
We also found that disclosure intimacy changed differently across mental health conditions. 
Anxiety, depression, and BPD showed shifts toward low-intimacy disclosures, primarily accompanied by reductions in moderate-intimacy content, while SI-related posts showed an increase in high-intimacy disclosures. 
However, when compared against the control corpus, only the shift from moderate- to low-intimacy disclosures for BPD remained statistically significant.

Together, these findings highlight the importance of examining not only the linguistic characteristics of mental health discourse, but also the depth of personal information individuals share during periods of platform change. Our work contributes a computational approach that combines psycholinguistic analysis with measures of disclosure intimacy to examine mental health self-disclosure surrounding a major platform transition. Our findings further motivate consideration of platform governance and moderation practices in designing online environments that support sensitive self-disclosure and mental health support-seeking.

\section{Related Work}\label{section:rw}

\subsection{Mental Health Self-Disclosure on Social Media}

Social media enables individuals to share personal experiences, seek support, and connect with others~\cite{saha2021life,saha2025life,de2014mental,berry2017whywetweetmh,luo2020self}.
Self-disclosure is shaped by individual motivations and social contexts~\cite{luo2020self,tang2012self}, with Social Penetration Theory distinguishing the breadth and depth of information shared~\cite{altman1973social,tang2012self,ernala2018characterizing}, while interpersonal intimacy also depends on perceived responsiveness~\cite{laurenceau1998intimacy}.
On social media, anonymity, context collapse, and imagined audiences shape self-disclosure decisions and intimacy ~\cite{ma2016anonymity,marwick2011tweet,marwick2014networked,litt2016imagined,jaidka2018facebook}.
These considerations are particularly relevant to mental health disclosures, where vulnerability and stigma can shape public sharing~\cite{de2014mental,budenz2022if,berry2017whywetweetmh,luo2020self}.
Research on Reddit and Twitter highlights the roles of anonymity, coping, social support, and connection in discussing mental health experiences~\cite{de2014mental,berry2017whywetweetmh,luo2020self}.
Beyond anticipated audiences, actual audience engagement has been examined in relation to subsequent disclosures of mental illness and their intimacy~\cite{ernala2018characterizing}.
Related studies investigate how peer responses and support relate to suicidal ideation risk and psychosocial outcomes in mental health communities, as well as changes following engagement with coping stories~\cite{de2017language,saha2020causal,yuan2023mental}.
More broadly, positive feedback has been associated with subsequent participation and content quality in online communities~\cite{lambert2025does}.
Together, this literature motivates examining the social contexts of mental health disclosure while distinguishing the occurrence of mental health discussion from the intimacy of the experiences shared~\cite{de2014mental,luo2020self,tang2012self,ernala2018characterizing}.

\subsection{Mental Health, Psycholinguistics, and Social Media}

Language use offers a basis for examining psychological processes, with Linguistic Inquiry and Word Count (LIWC) providing measures of affective, social, and cognitive dimensions of expression~\cite{pennebaker2003psychological,tausczik2010psychological,pennebaker2015development}.
Computational research has used social media language to study mental health expressions, including depression~\cite{mendu2020framework,de2013predicting,coppersmith2014quantifying}, stress~\cite{lin2014user}, suicidal ideation~\cite{de2016discovering,burnap2015machine}, and psychiatric medication-related mental health outcomes~\cite{saha2019social}.
Related work examined written-communication characteristics in mental health communities and linguistic markers associated with therapeutic outcomes following schizophrenia disclosures~\cite{park2018harnessing,ernala2017linguistic}.

Linguistic and psychosocial expressions can also change around major external events: prior studies examine language following the September 11 attacks~\cite{cohn2004linguistic}, stress expression following campus gun violence~\cite{saha2017modeling}, and psychosocial changes during the COVID-19 pandemic~\cite{saha2020psychosocial}.
Other research combined linguistic and behavioral measures to study psychosocial changes following engagement with mental health coping stories on Twitter~\cite{yuan2023mental}, news feeds on Bluesky~\cite{pal2026hidden}, and AI companion use~\cite{yuan2026mental}.
Although these approaches support population-level analyses of psychological expression, their construct validity and methodological limitations warrant caution~\cite{mendu2020framework,saha2022social,ernala2019methodological}.
Building on this work, we examine symptomatic and psycholinguistic expressions around Twitter's acquisition alongside disclosure intimacy, distinguishing changes in language from changes in how personally experiences are shared~\cite{tausczik2010psychological,saha2019social,tang2012self,ernala2018characterizing,ernala2019methodological,saha2022social}.

\subsection{Platform Governance and Changes in Online Participation}

Platform governance shapes online participation through moderation practices, interaction features, and community rules~\cite{wise2006moderation,fiesler2018reddit,chandrasekharan2017you}.
Research has examined the behavioral consequences of community-level moderation~\cite{chandrasekharan2017you,chandrasekharan2022quarantined}, the effects of witnessing post-removal explanations on bystanders~\cite{jhaver2024bystanders}, and patterns associated with Twitter account suspension~\cite{chowdhury2021examining}.
Beyond moderation, audits of algorithmic curation document relationships between feed ranking, content visibility, and engagement~\cite{chan2026examining}.
Broader policy changes can also disrupt established practices, as studies of Tumblr's 2018 adult-content ban document contested governance, migration and non-use, and implications for LGBTQ expression and visibility~\cite{sybert2022demise,edwards2021migration,southerton2021restricted}.

Elon Musk's acquisition marked a substantial transition in Twitter's governance and direction~\cite{murthy2024sociology,dupuis2025x,hickey2023auditing}.
Research surrounding the acquisition examines hate speech and inauthentic activity~\cite{hickey2023auditing,hickey2025x}, participation among environmental users and academics~\cite{chang2023environmental,bisbee2025vibes}, public sentiment~\cite{schmidt2023polarity}, and migration to alternatives such as Mastodon~\cite{cava2023drivers,zia2023flocking,jeong2024exploring}.
Together, these studies characterize changes in platform content, participation, and community composition~\cite{murthy2024sociology,dupuis2025x}.

Less is known about how sensitive first-person expression changed around the acquisition.
Although \citeauthor{mo2023analyzing} investigate schizophrenia stigmatization before and after the acquisition~\cite{mo2023analyzing}, stigmatizing discourse differs from disclosure of one's own mental health experiences~\cite{de2014mental,ernala2018characterizing}.
We therefore examine changes in the nature and intimacy of personal mental health disclosures around Twitter's acquisition~\cite{tang2012self,ernala2018characterizing}.

\section{Overview of Study Design and Data \label{section:data}}

\begin{figure}[t]
    \centering
    \includegraphics[width=0.5\textwidth]{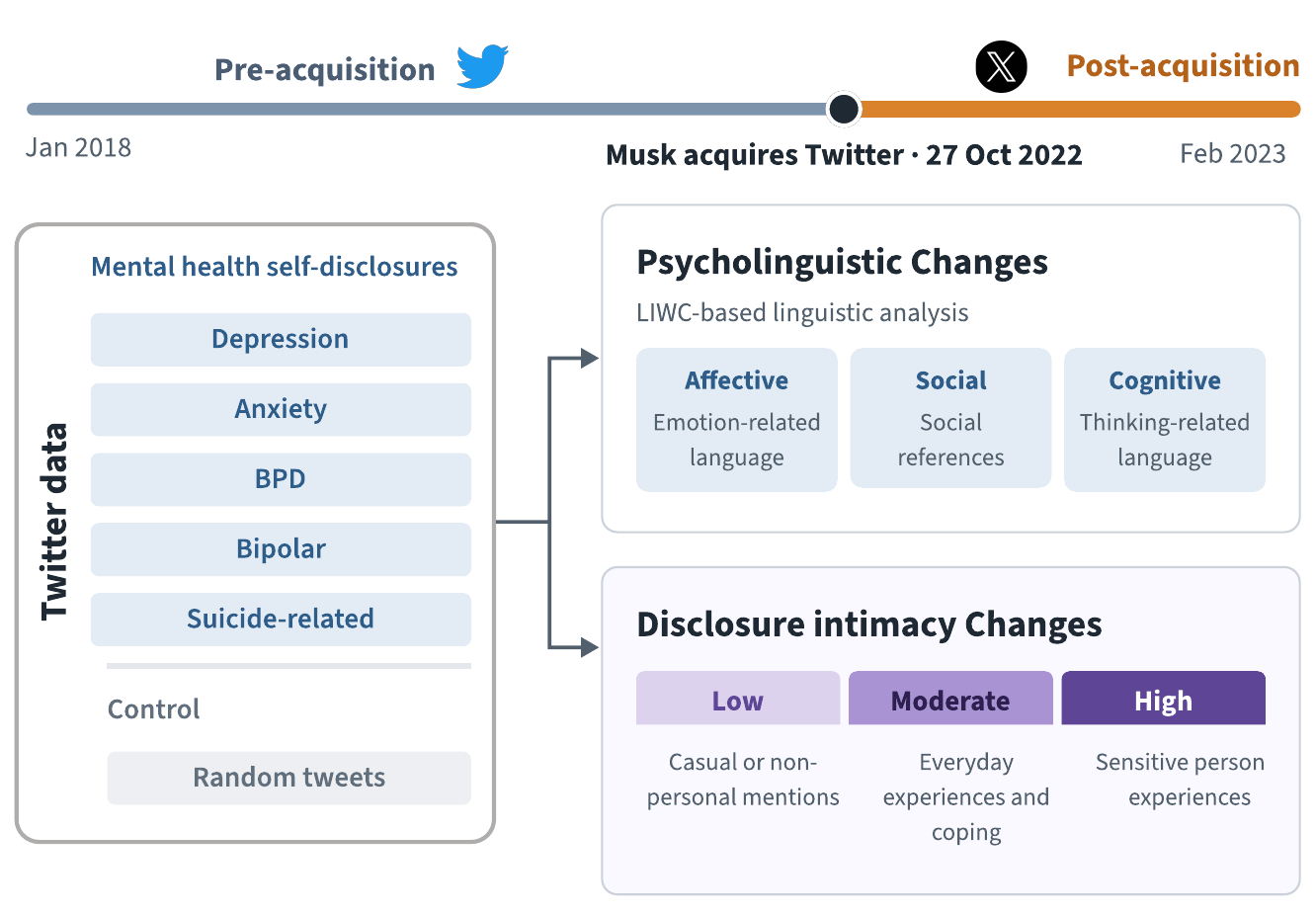}
    \caption{Overview of the study design.}
    \label{fig:study_overview}
\end{figure}

Figure~\ref{fig:study_overview} presents an overview of our study design. We examine mental health self-disclosures on Twitter between January 2018 and February 2023, using Elon Musk's acquisition on October 27, 2022, as the temporal division. Our analysis focuses on five mental health conditions and a comparison corpus of general Twitter posts. We characterize changes in mental health discourse along two dimensions: psycholinguistic attributes measured using LIWC and disclosure intimacy classified into low, moderate, and high levels.

\subsection{Twitter Data}
\label{subsec:twitter_data}

To examine the intimacy and linguistic characteristics of mental health disclosures, we collected Twitter data concerning depression, BD, anxiety, BPD, and SI-related experiences. Our collection focused on English-language Twitter posts posted between 1 January 2018 and 1 February 2023.

We began by searching for Twitter posts in which users explicitly referred to their own mental health diagnoses or experiences. Similar to prior study~\cite{saha2017inferring}, we searched for 40 phrases using the Twitter API, as presented in Table~\ref{tab:twitter_search_phrases}. For depression, BD, and anxiety, we used five phrases per condition, covering statements of having a condition, having received a diagnosis, or suffering from the condition. For borderline personality disorder, we used ten phrases covering both the full condition name and the abbreviation ``BPD''. The remaining fifteen phrases concerned suicidal ideation, desire, intent, or attempts, with present- and past-tense formulations capturing both current expressions and retrospective accounts.

This search strategy focused on potential first-person disclosures rather than general mentions of mental health. The resulting data reflect users' self-reported experiences; the search phrases alone do not verify clinical diagnoses or establish that every matching tweet constitutes a literal disclosure.

\subsection{Compiling Treatment and Control Data}
\label{subsec:treatment_control_data}

We compiled the Twitter posts retrieved through the disclosure queries into five samples, corresponding to depression, BD, anxiety, BPD, and SI-related experiences. 
The samples are not mutually exclusive, as posts may reference multiple conditions. 
This corresponds to our \textit{Treatment} dataset.
Table~\ref{tab:twitter_descriptive_statistics} presents the descriptive statistics.

We also collected an independent comparison (\textit{Control}) sample comprising 1,000 Twitter posts per week between 1 January 2018 and 1 February 2023. 
Collecting Twitter posts throughout the same observation period provided a temporal reference for examining whether linguistic changes also appeared in the Control corpus. 

\begin{table}[t]
\centering
\footnotesize
\caption{Descriptive statistics of the Twitter mental health dataset.}
\label{tab:twitter_descriptive_statistics}
\setlength{\tabcolsep}{3pt}
\resizebox{\columnwidth}{!}{
\begin{tabular}{lrrr}
\textbf{Dataset}
& \shortstack{\textbf{N.Users}}
& \shortstack{\textbf{N.Twitter posts}}
& \shortstack{\textbf{Twitter posts/User}\\\textbf{Mean (SD)}} \\
\toprule
Depression      & 142,356 & 175,503 & 1.23 (7.31) \\
\rowcollight Anxiety         & 304,393 & 350,101 & 1.15 (2.59) \\
BD         & 26,794  & 39,194  & 1.46 (1.96) \\
\rowcollight BPD      & 34,740  & 52,171  & 1.50 (4.37) \\
SI-related & 167,558 & 206,857 & 1.23 (4.67) \\
\rowcollight Control         & 248,374 & 266,364 & 1.07 (0.88) \\
\bottomrule
\end{tabular}}

\end{table}

\section{Methods} \label{sec:methods}

\subsection{Quantifying Psycholinguistic Changes in Mental Health Discourse}

\subsubsection{Psycholinguistic Analysis}

To characterize changes in the language of mental health disclosures following Twitter's acquisition, we used Linguistic Inquiry and Word Count (LIWC2015), a lexicon-based text analysis tool that measures linguistic and psychological characteristics of written language~\cite{pennebaker2015development,tausczik2010psychological}. Prior research has used psycholinguistic features to study mental health discourse and personal experiences shared on social media~\cite{de2014mental,saha2019computational}.

We applied LIWC to Twitter posts across the five mental health samples, considering the periods before and after the acquisition. The treated corpus comprised Twitter posts containing mental health disclosures, while the control corpus comprised comparison Twitter posts. Each tweet was analyzed individually, producing scores corresponding to the proportion of words associated with different LIWC categories.

We examined categories spanning affective attributes, cognitive processes, perception, interpersonal focus, temporal references, lexical and grammatical characteristics, biological concerns, personal and social concerns, and informal language. These included measures of emotional expression, cognitive processing, pronoun use, social references, and temporal orientation.

\subsubsection{Measuring Psycholinguistic Changes}

For each LIWC category, we calculated the mean scores separately for the treated and control corpora before and after the acquisition. We quantified the relative percentage change as $\Delta_{g,c}=100*(\overline{L}_{g,c,\mathrm{post}}/\overline{L}_{g,c,\mathrm{pre}}-1)$, where $\overline{L}_{g,c,\mathrm{pre}}$ and $\overline{L}_{g,c,\mathrm{post}}$ denote the mean LIWC scores for category $c$ in group $g$ before and after the acquisition, respectively.

We used $t$-tests to compare category-level scores between the two periods within each group, applying Bonferroni correction for multiple comparisons. We report the pre- and post-acquisition means, relative percentage changes, and adjusted statistical significance for each LIWC category.

\subsection{Operationalizing Intimacy of Self-Disclosures}

\subsubsection{Manual Annotation}

To characterize how personal mental health experiences were communicated, we adapted \citeauthor{ernala2018characterizing}'s approach to assess disclosure intimacy. The approach draws on Social Penetration Theory~\cite{altman1973social}, which distinguishes disclosures according to how much personal information individuals reveal to others. We applied this perspective across the five mental health samples, assigning individual Twitter posts to one of three ordered categories:

Low-intimacy disclosures contained brief or minimally elaborated references to mental health, including statements identifying a condition without further personal context. Moderate-intimacy disclosures described personal feelings, symptoms, or everyday difficulties with some contextual detail. High-intimacy disclosures contained detailed accounts of private experiences, vulnerabilities, or the personal consequences of mental health difficulties. References to a particular diagnosis or to SI-related did not, by themselves, determine the intimacy category.

To apply these categories, the authors first examined a subset of the corpus and developed annotation guidelines suited to the study's mental health samples. These guidelines were organized into a codebook containing definitions, distinguishing criteria, and representative examples for each intimacy level. Two annotators independently reviewed and assigned intimacy labels to individual Twitter posts based on the information disclosed in the text. Inter-annotator agreement was substantial (Cohen's $\kappa = 0.7323$). Disagreements were subsequently resolved by a third researcher to establish the final reference labels.

The resulting annotations served as reference labels for the automated classification of disclosure intimacy across the corpus.

\subsubsection{Automated Classification}

We used Qwen3-8B~\cite{yang2025qwen3} to classify the intimacy of individual mental health disclosures. The classification procedure followed an in-context learning approach, in which the model was provided with the annotation codebook and manually labeled reference examples. The codebook specified the definitions and criteria for low-, moderate-, and high-intimacy disclosures.
For each tweet, we selected 20 annotated examples from a reference set of 300 manually labeled Twitter posts using term frequency-inverse document frequency (TF-IDF) similarity. These examples were included in the classification prompt alongside the category definitions and instructions. The model then assigned one of the three intimacy labels based on the information disclosed in the tweet, without updating its parameters.
We evaluated the classifier on 40 held-out, manually annotated posts excluded from the retrieval pool. The classifier achieved 90.0\% accuracy and a macro-F1 score of 0.897. All four misclassifications involved low-intimacy posts classified as moderate, with all moderate- and high-intimacy posts correctly classified. We subsequently applied the classifier to the full corpus.

\subsubsection{Pre- and Post-Acquisition Analysis of Disclosure Intimacy}

We examined changes in disclosure intimacy across the five mental health conditions before and after Twitter's acquisition. For each condition, we calculated the proportion of posts classified as Low, Moderate, or High intimacy in each period and measured the change as the post-acquisition proportion minus the pre-acquisition proportion, expressed in percentage points.

We then compared these changes with the control corpus by subtracting the corresponding change in control posts from the change observed for each condition and intimacy category. This difference captures how much more or less each category changed relative to broader Twitter discourse over the same period. We interpret these comparisons descriptively rather than as causal treatment effects.

To assess whether temporal changes differed significantly from control, we fitted multinomial logistic regression models with intimacy category as the outcome and period, condition, and their interaction as predictors. Low intimacy served as the reference category. Standard errors were clustered by user to account for multiple posts from the same account, and Benjamini--Hochberg false discovery rate correction was applied across the five condition-control comparisons.

\section{Results}
\subsection{RQ1: Language Shifts}

Table~\ref{tab:liwc_comparison} summarizes pre- and post-acquisition changes in psycholinguistic expression among mental health disclosure (treated) and control Twitter posts. 
We observe notable changes in affective, cognitive, interpersonal, and social language, with several patterns differing between the two corpora.

\para{Affective and Cognitive Language.}
Treated posts showed significant declines in overall affect (16.65\%), anxiety (24.43\%), and sadness (18.26\%), while the corresponding changes in control posts were not significant. Positive emotion also declined, while anger showed no statistically significant change. Cognitive language exhibited mixed patterns: overall cognitive-process language increased slightly (1.89\%), while causal, tentative, certainty, and insight-related expressions declined. Notably, discrepancy-related language increased by 30.32\%, compared with 6.66\% in control posts.

\para{Interpersonal and Temporal Language.}
First-person singular pronouns increased by 10.52\% among treated posts but declined by 7.39\% in control posts. In contrast, references to ``we'' and ``you'' decreased among treated posts. Future-focused language increased substantially (42.75\%), while past-focused language declined (5.25\%), with control posts showing the opposite directions of change.

\para{Social and Other Linguistic Attributes.}
Social language declined by 11.79\% among treated posts, compared with a 1.17\% increase in control posts. Reductions were also observed in references to family, friends, work, affiliation, achievement, and power. Among other linguistic attributes, conjunctions and auxiliary verbs declined, while prepositions and verbs increased. Biological and health-related language showed smaller declines, whereas informal language remained relatively stable.

\subsection{RQ2: Intimacy Shifts}

\begin{table*}[t!]
\centering
\small
\caption{Changes in disclosure intimacy before and after the acquisition.
Significance is based on FDR-adjusted $p$-values
(* $p<.05$, ** $p<.01$, *** $p<.001$).}
\label{tab:rq2_prepost}
\resizebox{\textwidth}{!}{
\begin{tabular}{lrrrrrrrrr@{}l}
\textbf{Condition} &
\textbf{Pre $N$} &
\textbf{Post $N$} &
\textbf{$\Delta$ Low} &
\textbf{$\Delta$ Moderate} &
\textbf{$\Delta$ High} &
\textbf{Mean Pre} &
\textbf{Mean Post} &
\textbf{$\chi^2$} &
\multicolumn{2}{c}{\textbf{Sig.}} \\
\toprule
Anxiety
& 334,211 & 15,719
& +3.36 & -3.13 & -0.22
& 1.681 & 1.645
& 82.41 & *** \\

\rowcollight BD
& 36,797 & 2,395
& +1.67 & -1.22 & -0.46
& 1.826 & 1.805
& 4.68 &  \\

BPD
& 46,313 & 5,844
& +13.55 & -13.38 & -0.17
& 1.723 & 1.586
& 440.63 & *** \\

\rowcollight Depression
& 168,209 & 7,247
& +5.14 & -4.94 & -0.20
& 1.611 & 1.558
& 75.26 & *** \\

SI-related
& 188,672 & 17,877
& -0.77 & -1.26 & +2.03
& 2.288 & 2.316
& 55.41 & *** \\

\rowcollight Control
& 252,000 & 14,000
& +0.05 & +0.00 & -0.05
& 1.014 & 1.013
& 1.90 &  \\
\bottomrule
\end{tabular}
}

\end{table*}

\begin{table}[t!]
\centering
\small
\caption{Changes in disclosure intimacy relative to control
(\* $p_{\mathrm{FDR}}<.05$, \*\* $p_{\mathrm{FDR}}<.01$,
\*\*\* $p_{\mathrm{FDR}}<.001$).}
\label{tab:rq2_control}
\resizebox{\columnwidth}{!}{
\begin{tabular}{lrrrr@{}l}
\textbf{Condition} &
\textbf{$\Delta\Delta$ Low} &
\textbf{$\Delta\Delta$ Mod.} &
\textbf{$\Delta\Delta$ High} &
\textbf{Mod. vs. Low RRR} &
\multicolumn{1}{c}{\textbf{Sig.}} \\
\toprule
Anxiety
& +3.31 & -3.14 & -0.17
& 0.871
&  \\

\rowcollight BD
& +1.62 & -1.22 & -0.40
& 0.915
&  \\

BPD
& +13.50 & -13.38 & -0.12
& 0.553
& *** \\

\rowcollight Depression
& +5.09 & -4.94 & -0.15
& 0.813
&  \\

SI-related
& -0.82 & -1.27 & +2.09
& 0.821
&  \\
\bottomrule
\end{tabular}
}

\end{table}

\para{Changes in Disclosure Intimacy.}
Table~\ref{tab:rq2_prepost} shows shifts toward Low-intimacy disclosures following Twitter's acquisition, particularly for BPD (+13.55 percentage points), depression (+5.14), and anxiety (+3.36), accompanied by corresponding declines in Moderate-intimacy disclosures. High-intimacy disclosures changed comparatively little for these conditions, suggesting that the shifts primarily involved fewer accounts of everyday personal experiences, such as coping, relationships, and treatment, rather than substantial reductions in highly sensitive disclosures. BD showed no significant change. In contrast, SI-related posts exhibited an increase in High-intimacy disclosures (+2.03 percentage points), while the control corpus remained relatively stable.

\para{Changes Relative to Control.}
Table~\ref{tab:rq2_control} compares these changes against broader Twitter discourse. Although anxiety and depression showed shifts toward lower intimacy, neither differed significantly from control. BPD was the only condition exhibiting a significant relative decrease in Moderate versus Low intimacy (RRR $=0.553$, $p_{\mathrm{FDR}}<.001$). BD and SI-related disclosures also showed no significant differences relative to control, despite the increase in High-intimacy SI-related posts. These findings suggest that changes in disclosure intimacy were not uniform across mental health conditions, with the strongest evidence of a differential shift observed for BPD.

\section{Discussion and Conclusion}\label{section:discussion}

Our study examines changes in public mental health self-disclosure surrounding Twitter’s acquisition, considering both the linguistic characteristics of posts and the intimacy of personal experiences shared. These findings provide an opportunity to understand how broader platform-level transitions relate to the expression and depth of sensitive disclosures. We discuss their implications for Social Penetration Theory and platform governance, followed by considerations for designing supportive environments for mental health disclosure and computational approaches to studying disclosure intimacy.


\para{Social Penetration Theory and Platform Governance.}

Social Penetration Theory provides a lens for interpreting our findings as changes in disclosure depth~\cite{altman1973social}.
From this perspective, sharing increasingly personal information depends partly on the rewards and costs individuals associate with disclosure.
Prior research connects disclosure intimacy to reciprocal exchange and audience expectations~\cite{jiang2013perception,ma2016anonymity,ernala2018characterizing}, while studies of perceived observation and imagined surveillance show how audience awareness shapes posting behavior and self-presentation~\citep{saha2024observer,duffy2019you}.
Together, these studies highlight that personal sharing depends on the conditions under which individuals expect their disclosures to be observed and received.
Our findings bring attention to the governance environment surrounding these processes: platform rules and moderation may influence whether individuals anticipate support, indifference, or hostility when sharing vulnerable experiences.

The significant relative decline in moderate- versus low-intimacy BPD disclosures suggests a shift toward less personally elaborated accounts of everyday experiences, rather than a broad reduction in highly sensitive disclosures. However, corresponding patterns for anxiety, depression, and SI-related posts were not statistically significant relative to control. Viewed through Social Penetration Theory, these findings suggest that disclosure depth may vary across mental health contexts during platform transitions. Our aggregate analysis does not establish whether individual users withdrew from intimate exchanges or experienced increased disclosure costs.

\para{Freedom of Expression and Supportive Moderation.}
Following Twitter's acquisition, the platform underwent several governance changes, including the reinstatement of previously suspended accounts and the dissolution of its Trust and Safety Council~\cite{reuters2022amnesty,dang2022twitter}. 
These developments raise an important distinction between permitting freedom of expression and fostering an environment where individuals feel comfortable sharing stigmatized experiences. 
This distinction also relates to \textit{chilling effects}, whereby anticipated negative responses may discourage users from sharing sensitive experiences, even when such expression is permitted~\cite{stoycheff2016under,sleeper2013post}. 
The shift toward lower-intimacy BPD disclosures is consistent with this possibility, although our study cannot establish whether users self-censored.

Prior research has established the role of supportive moderation, community norms, and nonjudgmental interactions in facilitating sensitive self-disclosure and peer support~\cite{andalibi2017sensitive,marshall2024understanding,matias2019preventing,seering2019moderator}. Our findings extend this literature beyond community-level moderation by examining mental health disclosure during a broader transition in platform governance. 

This broader perspective highlights the need to consider how platform-wide rules and moderation practices interact with community norms in shaping the context of sensitive disclosures. However, our analysis does not isolate the effects of specific governance decisions.

\para{Design Implications.}

Our findings underscore the need for platform designs that support users in sharing intimate personal experiences related to mental health conditions and seeking social support.


Even moderately intimate accounts of coping, relationships, everyday functioning, and treatment may contain meaningful personal context that users wish to share selectively. Designing for these everyday disclosures therefore involves more than preventing harmful content; it also requires considering how individuals anticipate and manage their audiences. Platforms should accommodate different disclosure preferences rather than assuming that all users seek the same level of openness.

In this context, platforms could incorporate granular audience controls, pseudonymous participation, and dedicated spaces for mental health discussions, allowing users greater control over the visibility and reach of their disclosures. Consistently enforced community guidelines and moderation mechanisms that address harassment and stigmatizing responses may further support vulnerable participation. Importantly, these considerations extend beyond dedicated mental health communities to general-purpose social platforms, where personal disclosures may reach heterogeneous and unintended audiences.


\para{Methodological Implications.} Our work contributes a computational framework for examining changes in mental health self-disclosure surrounding major platform transitions. Rather than treating disclosure as a binary measure of whether personal information is shared, our approach distinguishes among low, moderate, and high levels of intimacy, enabling comparisons of disclosure depth across mental health conditions and observation periods. Integrating this measure with LIWC-based psycholinguistic analysis further captures changes in how these experiences are communicated. For instance, increased first-person language alongside shifts toward lower-intimacy disclosure illustrates how linguistic characteristics and disclosure depth can exhibit different patterns of change.

Additionally, incorporating a control corpus provides a baseline for examining whether changes in mental health disclosures differ from broader patterns of communication on the platform. 
Beyond Twitter's acquisition, this framework can inform future computational investigations of how platform-level changes relate to the expression of stigmatized experiences across online environments.
\section*{Limitations}

Our study has several limitations that offer directions for future research. To begin with, although Social Penetration Theory guides our analysis, we measure disclosure depth rather than reciprocal relationship development.
We do not observe perceived disclosure rewards and costs, audience responsiveness, or changes in interpersonal closeness.
Consequently, aggregate shifts toward lower intimacy cannot establish individual-level social depenetration.
Future research could combine longitudinal disclosure trajectories, audience interactions, and interviews to examine these processes.

Also, our analysis focuses on aggregate changes in mental health disclosures rather than individual-level trajectories. We examine conversations on Twitter before and after the acquisition but do not distinguish between users who continued participating, reduced their activity, or left the platform. Consequently, the observed changes may reflect shifts in how existing users disclose their experiences, changes in the composition of users contributing to these conversations, or both. Longitudinal analyses following individual users could provide a more nuanced understanding of how disclosure behaviors evolve in response to platform-level transitions.

Further, our study is subject to self-selection bias, as the dataset captures individuals who chose to publicly discuss mental health experiences on Twitter. Those who refrain from disclosing, engage in private conversations, or move to other platforms are not represented in our analysis. Moreover, the decision to disclose publicly may have changed after the acquisition, potentially shaping which experiences appear in our dataset. Further, individuals may disclose personal experiences for reasons beyond seeking social support, including establishing trust, managing impressions, or seeking social validation. Our analysis does not disentangle these motivations, which may influence both the content and intimacy of disclosures.

Again, our data were collected through Twitter's API while academic access was available. Consequently, our dataset reflects posts accessible through the API rather than an exhaustive record of mental health disclosures. Subsequent restrictions on academic API access also present challenges for reproducing and extending such longitudinal analyses. Additionally, deleted posts, suspended accounts, and other changes in content availability may affect the representation of historical discourse.

Our keyword-based retrieval strategy may influence the psycholinguistic characteristics observed in the data. Since retrieval phrases themselves contain words associated with LIWC categories, they may contribute directly to category-level scores. Changes in the relative prevalence of mental health conditions and retrieval phrases across periods may also contribute to aggregate linguistic differences. Furthermore, the retrieved posts represent potential self-disclosures rather than verified clinical experiences, and some expressions, particularly SI-related phrases, may be used figuratively.

Our comparison corpus also presents limitations. While general Twitter posts provide a reference for broader linguistic and disclosure patterns, they are not necessarily comparable to personal mental health disclosures. In particular, the concentration of low-intimacy posts in the control corpus limits its ability to capture changes in disclosure depth. Future work could examine comparable disclosures involving other personal or health-related experiences to provide a more informative baseline.

Finally, our post-acquisition observation period is relatively limited, capturing changes during the initial stages of Twitter's transition, while the pre-acquisition period spans several years. Consequently, broader temporal trends, seasonality, and contemporaneous events may contribute to the observed differences. Disclosure practices may also continue to evolve as users adapt to changes in platform governance, moderation, and community dynamics. Future research could examine longer and comparable observation periods to understand whether the observed patterns persist, intensify, or diminish over time.
\section*{Ethical Considerations}

This research examines changes in mental health self-disclosure on social media using computational and data-driven approaches. Given the sensitivity of mental health experiences and the personal nature of these disclosures, our work raises important ethical considerations surrounding privacy, interpretation, and potential misuse. Our study analyzes previously collected, publicly available Twitter posts without directly interacting with the individuals whose posts were analyzed. No institutional ethics review board approval or formal exemption determination was obtained for this research. Although these posts were publicly accessible, individuals may not anticipate their personal experiences being analyzed at scale.
Further, computational methods used to characterize disclosure intimacy could be misappropriated for surveillance, profiling, or other purposes that may harm vulnerable individuals. We emphasize that disclosure intimacy does not indicate mental health severity or clinical status, and our findings should not be used to draw conclusions about individual users.
Overall, this work is intended to advance understanding of mental health self-disclosure in changing social media environments, while encouraging responsible analysis and contextualized interpretation of sensitive online expressions.
\section*{AI Involvement Disclosure}
AI-assisted language editing (e.g., Grammarly, ChatGPT, Claude) was used exclusively to improve grammar and readability. The study design, analyses, interpretations, and experiments were conducted fully by the authors. 

\bibliography{references}

\appendix
\section{Appendix}\label{sec:appendix}

Table~\ref{tab:twitter_search_phrases} lists the 40 search phrases used to retrieve candidate self-disclosures for the five mental health samples, along with the collection procedure for the Control sample, which was not
keyword-filtered.

Table~\ref{tab:liwc_comparison} reports the full pre- and post-acquisition LIWC comparisons summarized in the Results section. For each category, it gives mean scores and relative changes for treated and control posts, the net change relative to control, and between-group $t$-statistics for each period.

Table~\ref{tab:intimacy_codebook} presents the codebook used for manual annotation and provided to the classifier. It includes definitions, criteria, and examples for each intimacy level, along with the decision rule for
assigning labels.

\setcounter{table}{0}
\setcounter{figure}{0}
\renewcommand{\thetable}{A\arabic{table}}
\renewcommand{\thefigure}{A\arabic{figure}}

\begin{table*}[t]
\centering
\caption{Search phrases and collection method for the five mental health disclosure samples and the control sample. The six samples are arranged in two panels for readability. All disclosure searches covered English-language tweets posted between 1 January 2018 and 1 February 2023.}
\label{tab:twitter_search_phrases}
\small
\setlength{\tabcolsep}{6pt}
\renewcommand{\arraystretch}{1.15}
\begin{tabular}{
@{}>{\raggedright\arraybackslash}p{\dimexpr(\textwidth-4\tabcolsep-2\arrayrulewidth)/3\relax}|
>{\raggedright\arraybackslash}p{\dimexpr(\textwidth-4\tabcolsep-2\arrayrulewidth)/3\relax}|
>{\raggedright\arraybackslash}p{\dimexpr(\textwidth-4\tabcolsep-2\arrayrulewidth)/3\relax}@{}
}
\hline
\textit{Depression} &
\textit{BD} &
\textit{Anxiety} \\
\hline

``I have depression''\par
``I have been diagnosed with depression''\par
``I was diagnosed with depression''\par
``I suffer from depression''\par
``I am suffering from depression''
&
``I have bipolar''\par
``I have been diagnosed with bipolar disorder''\par
``I was diagnosed with bipolar disorder''\par
``I suffer from bipolar''\par
``I am suffering from bipolar''
&
``I have anxiety''\par
``I have been diagnosed with anxiety''\par
``I was diagnosed with anxiety''\par
``I suffer from anxiety''\par
``I am suffering from anxiety''
\\
\hline
\textit{BPD} &
\textit{SI-related} &
\textit{Control} \\
\hline

``I have borderline personality disorder''\par
``I have been diagnosed with borderline personality disorder''\par
``I was diagnosed with borderline personality disorder''\par
``I suffer from borderline personality disorder''\par
``I am suffering from borderline personality disorder''\par
``I have bpd''\par
``I have been diagnosed with bpd''\par
``I was diagnosed with bpd''\par
``I suffer from bpd''\par
``I am suffering from bpd''
&
``I have suicidal ideation''\par
``I want to kill myself''\par
``I wanted to kill myself''\par
``I want to suicide''\par
``I wanted to suicide''\par
``I am going to kill myself''\par
``I was going to kill myself''\par
``I tried to kill myself''\par
``I try to kill myself''\par
``I tried to suicide''\par
``I try to suicide''\par
``I want to commit suicide''\par
``I wanted to commit suicide''\par
``I am going to commit suicide''\par
``I was going to commit suicide''
&
Twitter API collection of 1,000 tweets per week.\par
Observation period: 1 January 2018--1 February 2023.
\\
\hline
\end{tabular}
\end{table*}

\begin{table*}[t]
\centering
\sffamily
\footnotesize
\caption{Pre- vs.\ post-acquisition LIWC comparisons for treated (mental health disclosure) and control posts. $\Delta\%$: within-group change. ATE\,$\%=\beta/\mu_{\mathrm{cf}}\times100$, where $\beta=\Delta_{Tr}-\Delta_{Ct}$ and $\mu_{\mathrm{cf}}=\mu_{Tr,\mathrm{pre}}+\Delta_{Ct}$ ($^{\dagger}$undefined, $\mu_{\mathrm{cf}}<0$). Pre/Post $t$: between-group tests. Bonferroni-adjusted $^{*}p<.05$, $^{**}p<.01$, $^{***}p<.001$.}
\label{tab:liwc_comparison}
\setlength{\tabcolsep}{3pt}

\begin{minipage}[t]{0.5\textwidth}
\vspace{0pt}
\centering
\resizebox{\linewidth}{!}{%
\begin{tabular}{lrrr@{}lrrr@{}lrr@{}lr@{}l}
\toprule
\textbf{LIWC}
& \multicolumn{4}{c}{\textbf{Treated (Tr.)}}
& \multicolumn{4}{c}{\textbf{Control (Ct.)}}
& \multicolumn{1}{c}{\textbf{ATE}}
& \multicolumn{4}{c}{\textbf{Tr. vs. Ct.}} \\
\cmidrule(lr){2-5}
\cmidrule(lr){6-9}
\cmidrule(lr){10-10}
\cmidrule(lr){11-14}
& \textbf{Pre} & \textbf{Post}
& \multicolumn{2}{c}{$\boldsymbol{\Delta\%}$}
& \textbf{Pre} & \textbf{Post}
& \multicolumn{2}{c}{$\boldsymbol{\Delta\%}$}
& $\boldsymbol{\%}$
& \multicolumn{2}{c}{\textbf{Pre} $\boldsymbol{t}$}
& \multicolumn{2}{c}{\textbf{Post} $\boldsymbol{t}$} \\
\toprule

\sechead{Affective Attributes}\\
Affect & 0.0669 & 0.0557 & -16.65 & $^{***}$
& 0.0537 & 0.0525 & -2.17 &
& -15.17 & $66.196$ & $^{***}$ & $3.828$ & $^{**}$ \\
Pos. Emotion & 0.0192 & 0.0173 & -10.28 & $^{***}$
& 0.0400 & 0.0388 & -2.86 &
& -4.62 & $-123.821$ & $^{***}$ & $-30.190$ & $^{***}$ \\
Neg. Emotion & 0.0001 & 0.0001 & -5.61 &
& 0.0009 & 0.0006 & -34.87 & $^{**}$
& n/a$^{\dagger}$ & $-34.313$ & $^{***}$ & $-6.593$ & $^{***}$ \\
\hdashline
Anger & 0.0038 & 0.0037 & -2.93 &
& 0.0039 & 0.0041 & +5.26 &
& -7.93 & $-2.130$ &  & $-1.780$ &  \\
Anxiety & 0.0321 & 0.0243 & -24.43 & $^{***}$
& 0.0010 & 0.0011 & +9.68 &
& -24.65 & $494.725$ & $^{***}$ & $96.052$ & $^{***}$ \\
Sadness & 0.0062 & 0.0050 & -18.26 & $^{***}$
& 0.0038 & 0.0036 & -5.53 &
& -15.37 & $41.843$ & $^{***}$ & $6.417$ & $^{***}$ \\
\hdashline
Swear & 0.0025 & 0.0025 & -2.96 &
& 0.0037 & 0.0033 & -10.18 &
& +13.84 & $-20.598$ & $^{***}$ & $-3.498$ & $^{*}$ \\

\sechead{Cognitive Attributes} \\
Cog. Proc. & 0.0969 & 0.0987 & +1.89 & $^{***}$
& 0.0640 & 0.0662 & +3.37 &
& -0.33 & $169.742$ & $^{***}$ & $39.015$ & $^{***}$ \\
Cause & 0.0148 & 0.0132 & -10.89 & $^{***}$
& 0.0087 & 0.0090 & +3.57 &
& -12.72 & $93.021$ & $^{***}$ & $15.038$ & $^{***}$ \\
Negation & 0.0152 & 0.0146 & -3.83 & $^{***}$
& 0.0127 & 0.0144 & +14.07 & $^{***}$
& -13.93 & $29.317$ & $^{***}$ & $0.399$ &  \\
\hdashline
Tentative & 0.0160 & 0.0150 & -6.75 & $^{***}$
& 0.0135 & 0.0137 & +1.64 &
& -8.02 & $31.806$ & $^{***}$ & $3.742$ & $^{*}$ \\
Certainty & 0.0102 & 0.0096 & -6.29 & $^{***}$
& 0.0110 & 0.0113 & +2.15 &
& -8.40 & $-8.955$ & $^{***}$ & $-4.351$ & $^{**}$ \\
Differentiation & 0.0256 & 0.0236 & -7.85 & $^{***}$
& 0.0177 & 0.0185 & +4.17 &
& -10.44 & $84.227$ & $^{***}$ & $13.039$ & $^{***}$ \\
\hdashline
Discrepancy & 0.0250 & 0.0325 & +30.32 & $^{***}$
& 0.0106 & 0.0114 & +6.66 &
& +26.72 & $176.051$ & $^{***}$ & $57.332$ & $^{***}$ \\

\sechead{Perception} \\
Perceptual & 0.0145 & 0.0142 & -2.21 &
& 0.0172 & 0.0172 & +0.27 &
& -2.52 & $-27.342$ & $^{***}$ & $-6.889$ & $^{***}$ \\
Feel & 0.0049 & 0.0042 & -13.58 & $^{***}$
& 0.0039 & 0.0039 & -0.20 &
& -13.44 & $18.282$ & $^{***}$ & $1.340$ &  \\ \hdashline
Hear & 0.0044 & 0.0042 & -5.83 & $^{*}$
& 0.0035 & 0.0035 & -0.99 &
& -5.08 & $20.835$ & $^{***}$ & $3.367$ &  \\
See & 0.0049 & 0.0055 & +12.10 & $^{***}$
& 0.0090 & 0.0091 & +1.44 &
& +9.19 & $-58.414$ & $^{***}$ & $-11.075$ & $^{***}$ \\
Insight & 0.0161 & 0.0152 & -5.34 & $^{***}$
& 0.0102 & 0.0104 & +2.16 &
& -6.62 & $80.909$ & $^{***}$ & $15.380$ & $^{***}$ \\

\sechead{Interpersonal Focus} \\
All Pronouns & 0.2120 & 0.2227 & +5.04 & $^{***}$
& 0.1163 & 0.1168 & +0.44 &
& +4.79 & $406.191$ & $^{***}$ & $106.375$ & $^{***}$ \\
Personal Pron. & 0.1767 & 0.1907 & +7.91 & $^{***}$
& 0.0798 & 0.0789 & -1.18 &
& +8.49 & $479.306$ & $^{***}$ & $131.811$ & $^{***}$ \\
Impersonal Pron. & 0.0353 & 0.0321 & -9.29 & $^{***}$
& 0.0365 & 0.0379 & +4.02 &
& -12.90 & $-8.153$ & $^{***}$ & $-9.918$ & $^{***}$ \\ \hdashline
I & 0.1549 & 0.1712 & +10.52 & $^{***}$
& 0.0400 & 0.0370 & -7.39 & $^{***}$
& +12.67 & $706.256$ & $^{***}$ & $196.333$ & $^{***}$ \\
We & 0.0019 & 0.0016 & -17.62 & $^{***}$
& 0.0056 & 0.0054 & -3.95 &
& -6.70 & $-72.432$ & $^{***}$ & $-18.024$ & $^{***}$ \\
\hdashline
You & 0.0125 & 0.0105 & -16.58 & $^{***}$
& 0.0215 & 0.0211 & -1.75 &
& -14.01 & $-81.952$ & $^{***}$ & $-23.981$ & $^{***}$ \\
She/He & 0.0041 & 0.0042 & +2.65 &
& 0.0084 & 0.0098 & +15.74 & $^{***}$
& -22.62 & $-65.468$ & $^{***}$ & $-18.543$ & $^{***}$ \\
They & 0.0032 & 0.0032 & -0.15 &
& 0.0043 & 0.0056 & +30.06 & $^{***}$
& -28.54 & $-23.437$ & $^{***}$ & $-10.836$ & $^{***}$ \\

\sechead{Temporal References} \\
Future Focus & 0.0100 & 0.0143 & +42.75 & $^{***}$
& 0.0091 & 0.0088 & -3.87 &
& +47.97 & $12.658$ & $^{***}$ & $18.551$ & $^{***}$ \\
Past Focus & 0.0252 & 0.0238 & -5.25 & $^{***}$
& 0.0215 & 0.0231 & +7.24 & $^{*}$
& -10.77 & $34.066$ & $^{***}$ & $1.668$ &  \\
Present Focus & 0.1537 & 0.1545 & +0.54 &
& 0.0846 & 0.0838 & -0.90 &
& +1.04 & $333.017$ & $^{***}$ & $81.340$ & $^{***}$ \\

\sechead{Biological Concerns} \\
Biological & 0.0133 & 0.0128 & -4.15 & $^{**}$
& 0.0134 & 0.0120 & -10.07 & $^{*}$
& +6.64 & $-0.646$ &  & $1.861$ &  \\
Body & 0.0032 & 0.0029 & -7.99 & $^{**}$
& 0.0035 & 0.0031 & -11.27 &
& +5.01 & $-6.700$ & $^{***}$ & $-0.948$ &  \\
Health & 0.0067 & 0.0064 & -5.01 & $^{*}$
& 0.0031 & 0.0030 & -2.87 &
& -3.74 & $77.339$ & $^{***}$ & $16.083$ & $^{***}$ \\ \hdashline
Ingestion & 0.0014 & 0.0014 & -0.68 &
& 0.0022 & 0.0017 & -22.87 & $^{*}$
& +56.64 & $-20.645$ & $^{***}$ & $-2.495$ &  \\
Death & 0.0008 & 0.0009 & +5.49 &
& 0.0008 & 0.0008 & -0.20 &
& +5.70 & $0.382$ &  & $0.624$ &  \\
Sexual & 0.0012 & 0.0013 & +11.47 &
& 0.0016 & 0.0017 & +6.18 &
& +2.90 & $-10.964$ & $^{***}$ & $-2.056$ &  \\

\bottomrule
\end{tabular}%
}
\end{minipage}\hfill
\begin{minipage}[t]{0.5\textwidth}
\vspace{0pt}
\centering
\resizebox{\linewidth}{!}{%
\begin{tabular}{lrrr@{}lrrr@{}lrr@{}lr@{}l}
\toprule
\textbf{LIWC}
& \multicolumn{4}{c}{\textbf{Treated (Tr.)}}
& \multicolumn{4}{c}{\textbf{Control (Ct.)}}
& \multicolumn{1}{c}{\textbf{ATE}}
& \multicolumn{4}{c}{\textbf{Tr. vs. Ct.}} \\
\cmidrule(lr){2-5}
\cmidrule(lr){6-9}
\cmidrule(lr){10-10}
\cmidrule(lr){11-14}
& \textbf{Pre} & \textbf{Post}
& \multicolumn{2}{c}{$\boldsymbol{\Delta\%}$}
& \textbf{Pre} & \textbf{Post}
& \multicolumn{2}{c}{$\boldsymbol{\Delta\%}$}
& $\boldsymbol{\%}$
& \multicolumn{2}{c}{\textbf{Pre} $\boldsymbol{t}$}
& \multicolumn{2}{c}{\textbf{Post} $\boldsymbol{t}$} \\
\toprule

\sechead{Lexical Density and Awareness} \\
Function Words & 0.5294 & 0.5300 & +0.11 &
& 0.3398 & 0.3456 & +1.69 &
& -0.97 & $463.980$ & $^{***}$ & $106.807$ & $^{***}$ \\
Article & 0.0264 & 0.0242 & -8.31 & $^{***}$
& 0.0358 & 0.0371 & +3.59 &
& -12.56 & $-80.688$ & $^{***}$ & $-26.120$ & $^{***}$ \\
Preposition & 0.0930 & 0.0970 & +4.27 & $^{***}$
& 0.0727 & 0.0738 & +1.44 &
& +3.11 & $120.147$ & $^{***}$ & $32.606$ & $^{***}$ \\
Conjunction & 0.0604 & 0.0549 & -9.21 & $^{***}$
& 0.0319 & 0.0325 & +2.04 &
& -10.18 & $235.438$ & $^{***}$ & $44.183$ & $^{***}$ \\
\hdashline

Adverb & 0.0513 & 0.0463 & -9.62 & $^{***}$
& 0.0410 & 0.0411 & +0.25 &
& -9.80 & $68.176$ & $^{***}$ & $8.160$ & $^{***}$ \\
Aux. Verb & 0.1090 & 0.1056 & -3.13 & $^{***}$
& 0.0575 & 0.0595 & +3.39 &
& -4.83 & $300.868$ & $^{***}$ & $63.942$ & $^{***}$ \\

\hdashline
Verb & 0.1895 & 0.1930 & +1.88 & $^{***}$
& 0.1183 & 0.1196 & +1.08 &
& +1.19 & $308.638$ & $^{***}$ & $75.613$ & $^{***}$ \\
Adjective & 0.0293 & 0.0279 & -4.98 & $^{***}$
& 0.0400 & 0.0398 & -0.37 &
& -4.49 & $-66.711$ & $^{***}$ & $-17.860$ & $^{***}$ \\
Compare & 0.0149 & 0.0142 & -4.57 & $^{***}$
& 0.0151 & 0.0150 & -1.04 &
& -3.54 & $-2.707$ &  & $-2.070$ &  \\

\hdashline
Interrogative & 0.0144 & 0.0130 & -9.52 & $^{***}$
& 0.0120 & 0.0120 & -0.06 &
& -9.48 & $29.257$ & $^{***}$ & $3.035$ &  \\
Number & 0.0037 & 0.0034 & -9.78 & $^{***}$
& 0.0046 & 0.0046 & -1.14 &
& -8.47 & $-18.992$ & $^{***}$ & $-6.331$ & $^{***}$ \\
Quantifier & 0.0122 & 0.0113 & -7.05 & $^{***}$
& 0.0136 & 0.0135 & -1.19 &
& -5.78 & $-18.726$ & $^{***}$ & $-6.500$ & $^{***}$ \\

\sechead{Personal and Social Concerns} \\
Social & 0.0462 & 0.0408 & -11.79 & $^{***}$
& 0.0649 & 0.0656 & +1.17 &
& -13.21 & $-95.897$ & $^{***}$ & $-30.942$ & $^{***}$ \\
Family & 0.0027 & 0.0024 & -9.09 & $^{**}$
& 0.0030 & 0.0024 & -19.05 & $^{*}$
& +15.11 & $-5.947$ & $^{***}$ & $0.249$ &  \\
Friend & 0.0014 & 0.0013 & -12.65 & $^{***}$
& 0.0016 & 0.0014 & -15.32 &
& +5.75 & $-6.000$ & $^{***}$ & $-1.061$ &  \\ \hdashline
Female & 0.0036 & 0.0032 & -13.45 & $^{***}$
& 0.0046 & 0.0042 & -9.74 &
& -1.26 & $-17.667$ & $^{***}$ & $-4.780$ & $^{***}$ \\
Male & 0.0042 & 0.0047 & +12.32 & $^{***}$
& 0.0095 & 0.0112 & +17.24 & $^{***}$
& -19.49 & $-71.766$ & $^{***}$ & $-18.067$ & $^{***}$ \\ \hdashline
Home & 0.0015 & 0.0013 & -15.79 & $^{***}$
& 0.0015 & 0.0014 & -7.83 &
& -8.78 & $1.074$ &  & $-0.884$ &  \\
Religion & 0.0011 & 0.0014 & +24.14 & $^{***}$
& 0.0024 & 0.0028 & +13.91 &
& -4.66 & $-26.280$ & $^{***}$ & $-6.246$ & $^{***}$ \\
Money & 0.0011 & 0.0010 & -7.78 &
& 0.0026 & 0.0033 & +24.59 &
& -42.55 & $-42.656$ & $^{***}$ & $-11.982$ & $^{***}$ \\ \hdashline
Work & 0.0040 & 0.0034 & -15.13 & $^{***}$
& 0.0045 & 0.0044 & -2.82 &
& -12.37 & $-9.695$ & $^{***}$ & $-4.851$ & $^{***}$ \\
Leisure & 0.0028 & 0.0026 & -4.84 &
& 0.0052 & 0.0047 & -8.65 &
& +13.66 & $-47.350$ & $^{***}$ & $-9.855$ & $^{***}$ \\
Drives & 0.0363 & 0.0328 & -9.78 & $^{***}$
& 0.0482 & 0.0474 & -1.54 &
& -7.89 & $-72.334$ & $^{***}$ & $-21.187$ & $^{***}$ \\ \hdashline
Achievement & 0.0075 & 0.0069 & -7.91 & $^{***}$
& 0.0083 & 0.0083 & +0.19 &
& -8.10 & $-12.922$ & $^{***}$ & $-5.428$ & $^{***}$ \\
Affiliation & 0.0091 & 0.0079 & -13.40 & $^{***}$
& 0.0150 & 0.0136 & -9.58 & $^{**}$
& +2.72 & $-61.806$ & $^{***}$ & $-15.238$ & $^{***}$ \\
Power & 0.0102 & 0.0089 & -13.06 & $^{***}$
& 0.0133 & 0.0136 & +2.36 &
& -15.64 & $-35.066$ & $^{***}$ & $-12.440$ & $^{***}$ \\
Reward & 0.0090 & 0.0082 & -9.36 & $^{***}$
& 0.0148 & 0.0150 & +1.77 &
& -11.90 & $-61.920$ & $^{***}$ & $-16.772$ & $^{***}$ \\ \hdashline
Risk & 0.0041 & 0.0042 & +3.09 &
& 0.0029 & 0.0030 & +3.06 &
& +0.85 & $25.640$ & $^{***}$ & $6.102$ & $^{***}$ \\

\sechead{Relativity} \\
Relativity & 0.0745 & 0.0735 & -1.34 &
& 0.0777 & 0.0761 & -2.06 &
& +0.83 & $-15.183$ & $^{***}$ & $-2.952$ &  \\
Motion & 0.0128 & 0.0164 & +28.12 & $^{***}$
& 0.0132 & 0.0125 & -5.62 &
& +36.01 & $-4.927$ & $^{***}$ & $11.122$ & $^{***}$ \\
Space & 0.0312 & 0.0289 & -7.28 & $^{***}$
& 0.0388 & 0.0397 & +2.32 &
& -9.88 & $-56.264$ & $^{***}$ & $-18.624$ & $^{***}$ \\ \hdashline
Time & 0.0314 & 0.0290 & -7.80 & $^{***}$
& 0.0273 & 0.0253 & -7.38 & $^{**}$
& -1.49 & $32.855$ & $^{***}$ & $7.386$ & $^{***}$ \\

\sechead{Informal Language} \\
Informal & 0.0191 & 0.0197 & +3.18 &
& 0.0285 & 0.0259 & -9.15 & $^{**}$
& +19.50 & $-60.739$ & $^{***}$ & $-9.462$ & $^{***}$ \\
Netspeak & 0.0098 & 0.0096 & -2.25 &
& 0.0157 & 0.0127 & -19.05 & $^{***}$
& +40.52 & $-54.655$ & $^{***}$ & $-7.728$ & $^{***}$ \\
Assent & 0.0040 & 0.0047 & +18.58 & $^{***}$
& 0.0059 & 0.0060 & +1.19 &
& +16.50 & $-24.824$ & $^{***}$ & $-3.598$ & $^{*}$ \\ \hdashline
Nonfluencies & 0.0021 & 0.0019 & -9.27 & $^{*}$
& 0.0033 & 0.0036 & +10.76 &
& -22.17 & $-21.694$ & $^{***}$ & $-6.624$ & $^{***}$ \\
Filler & 0.0006 & 0.0006 & +9.02 &
& 0.0004 & 0.0004 & +8.52 &
& +3.18 & $9.745$ & $^{***}$ & $2.064$ &  \\

\bottomrule
\end{tabular}%
}
\end{minipage}

\end{table*}

\begin{table*}[t]
\centering
\footnotesize
\caption{Codebook for disclosure intimacy classification.}
\label{tab:intimacy_codebook}
\setlength{\tabcolsep}{5pt}

\begin{tabular}{
@{}p{0.11\textwidth}
p{0.39\textwidth}
p{0.44\textwidth}@{}
}
\toprule
\textbf{Level} &
\textbf{Definition and Criteria} &
\textbf{Examples} \\
\midrule

\textbf{Low (1)}
&
Little or no substantive disclosure of the author's lived mental health experience. This includes bare diagnoses, lists of diagnoses, background or incidental mentions of a condition, general discussion of mental health, third-party information, ordinary situational worry, casual conversation, and figurative or humorous uses of mental health language. A mental health keyword alone is not sufficient for Moderate or High intimacy.
&
``I have depression.'' $\rightarrow$ Low \par
``I have anxiety and ADHD.'' $\rightarrow$ Low \par
``I think I might have BPD.'' $\rightarrow$ Low \par
``I have depression and I love playing video games.'' $\rightarrow$ Low \par
``That actor is so attractive I wanna die.'' $\rightarrow$ Low
\\
\midrule

\textbf{Moderate (2)}
&
A personal disclosure describing how a mental health condition affects everyday life. This includes effects on work or school, relationships, sleep or eating, coping strategies, medication use, disclosure to family or friends, social interactions, and other functional or behavioral consequences. The post must describe a concrete lived experience rather than only naming the condition.
&
``My anxiety makes it difficult to concentrate at work.'' $\rightarrow$ Moderate \par
``Because of my depression, I sometimes struggle to get out of bed for class.'' $\rightarrow$ Moderate \par
``Watching streams helps me cope with my depression.'' $\rightarrow$ Moderate \par
``I take medication for my anxiety every morning.'' $\rightarrow$ Moderate \par
``My BPD sometimes makes me jealous in relationships.'' $\rightarrow$ Moderate
\\
\midrule

\textbf{High (3)}
&
Disclosure of highly sensitive lived mental health experiences. This includes suicidal ideation or suicide attempts, self-harm, mania, psychosis or hallucinations, dissociation, other sensitive symptoms, personal experiences of stigma, and seeking emotional or social support around a mental health condition. High-intimacy information takes precedence when a post also contains Moderate-level information.
&
``I want to kill myself.'' $\rightarrow$ High \par
``I attempted suicide when I was younger.'' $\rightarrow$ High \par
``I'm hearing voices again and I'm scared.'' $\rightarrow$ High \par
``I've been experiencing manic episodes.'' $\rightarrow$ High \par
``I keep dissociating and losing track of what's happening.'' $\rightarrow$ High
\\
\midrule

\multicolumn{3}{@{}p{0.96\textwidth}@{}}{
\textbf{Decision rule.}
Posts were interpreted in their full context. We first distinguished personal experience from jokes, quotations, general discussion, or third-party information. Highly sensitive personal disclosures were assigned High; otherwise, concrete everyday experiences related to functioning, relationships, coping, medication, or condition management were assigned Moderate. Posts containing only diagnostic labels, incidental mentions, casual discussion, figurative language, or general opinions were assigned Low.
}
\\

\bottomrule
\end{tabular}
\end{table*}

\end{document}